\documentclass[11pt]{article}
\usepackage{moriond,epsfig}

\def\Journal#1#2#3#4{{#1} {\bf #2}, #3 (#4)}

\def\NIM{\em Nucl. Instrum. Methods}

\def\NPB{{\em Nucl. Phys.} B}

\def\PRD{{\em Phys. Rev.} D}

\def\EPJC{{\em . Eur.Phys.J.} C}
\def\CPC{\em Computer Physics Commun.}

\def\be{\begin{equation}}
\def\ee{\end{equation}}
\def\bea{\begin{eqnarray}}
\def\eea{\end{eqnarray}}

\def\pim{\pi^-}
\def\pip{\pi^+}
\def\km{K^-}
\def\kp{K^+}

\begin{document}
\vspace*{4cm}
\title{Hadron Multiplicities at HERMES}

\author{A.~Hillenbrand and M.~Hartig for the HERMES
Collaboration}

\address{DESY, 22603 Hamburg, Germany}

\maketitle\abstracts{
Hadron multiplicities of $\pim$, $\pip$, $\km$ and $\kp$ have been
measured in the deep-inelastic scattering of 27.5~GeV
positrons off a hydrogen target. The data used in this analysis 
have been collected during the 2000 HERA running period. 
The multiplicities were obtained for 0.15$< z <$0.9 
for $<Q^2>$ = 2.5~GeV$^2$.}

\section{Introduction}

\begin{figure}
\centering
\psfig{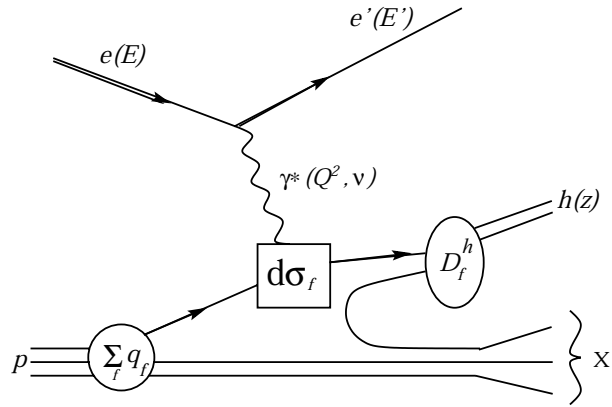}
\caption{Semi-inclusive hadron electroproduction diagram.
\label{fig:sidis}}
\centering
\end{figure}

The semi-inclusive production of mesons in Deep-Inelastic Scattering
(DIS) is a good tool to test the quark-parton model and QCD. 
A schematic diagram and the relevant variables for the process 
are shown in Figure~\ref{fig:sidis}. In the proton rest frame, 
the energy of the exchanged virtual photon $\gamma^*$ is $\nu = E - E'$ 
($E$ and $E'$ being the energies of the incident and scattered positrons
respectively), while its negative  squared four-momentum is $Q^2$. 
The quantity  $x = Q^2/2M\nu$, where M is the proton mass, is 
(in the Breit reference frame)
the fraction of the momentum of the nucleon carried by the struck quark. 

The parton distribution function $q_f$ describes the momentum distribution
of quarks in the nucleon, while the fragmentation function $D^h_f$
is a measure of the probability that a quark of flavor $f$ fragments
into a hadron of energy $E_h = z\nu$. The quantity $d\sigma_f$ is the 
cross-section for the absorption of the virtual photon by the struck
quark. The quantity of interest in this paper is the hadron
differential multiplicity, or the number $N^h$ of hadrons produced in
DIS, normalized to the total number ($N_{DIS}$) of inclusive DIS
events. In the QCD improved quark-parton model, it is given by
the expression: 
\begin{equation}
\mathcal{M}^h = \frac{1}{N_{\mathrm{DIS}}(Q^2)}
                \cdot 
                \frac{dN^h(z,Q^2)}{dz}
	  = \frac{
	    \sum_f e_f^2 \int_0^1 dx\, q_f(x,Q^2) D^{h}_f(z,Q^2)
	    }{
            \sum_f e_f^2 \int_0^1 dx\, q_f({x},Q^2)
	    },
\end{equation}
where the sum is over quarks and anti-quarks of flavor $f$, and $e_f$ 
is the quark charge in units of the elementary charge. 

The measurements described here were performed with the HERMES
experiment using the 27.5~GeV positron beam stored in the HERA ring
at DESY.
The HERMES detector is an open forward spectrometer consisting of two
identical halves above and beneath the HERA beam pipe. It has three
main components: the spectrometer magnet, the tracking system
consisting of three sets of tracking chambers in front of, inside and
behind the spectrometer magnet, and a particle identification system.
%The aperture of the magnet  limits the geometrical acceptance of the
%spectrometer to $\pm$ 140~mrad in the vertical and 
The particle identification (PID) system, which is essential for this
analysis, consists of four detectors: a lead glass calorimeter, a
preshower detector, a transition radiation detector (TRD) 
and a ring imaging Cherenkov counter (RICH).
A detailed description of the detector components can be found in
reference~\cite{hermes,rich}.
The data used in this analysis were collected during the 2000 HERA
running period. The hydrogen gas target internal to the positron
storage ring was operated in unpolarized mode.

\section{Data Analysis}

The identification of the scattered positron was accomplished using
a probability algorithm utilizing the response of the four PID
detectors. This system provided positron identification with an
average efficiency of 98~\% and a hadron contamination of less
than 1~\%. Events were selected by imposing the kinematic restrictions
$Q^2 \ge 1$~GeV$^2$ and $y \le 0.85$, where $ y = \nu/E$ is the virtual
photon fractional energy.
In order to exclude effects from nucleon resonances as well as
kinematic regions with inadequate geometrical acceptance, the
additional requirement $W^2 \ge 10$~GeV$^2$ was imposed for this analysis.  
The data cover the region of 0.15 $< z <$ 0.9. 

The RICH detector~\cite{rich} identifies pions, kaons, and protons
in a momentum range 2~GeV $< p <$ 15~GeV. 
The performance of the RICH system is summarized by a matrix $P$ with 
elements $P^i_t$, being the probabilities for identifying the true particle 
type $t$ as type $i$. The off-diagonal elements of $P$, i.e. the 
misidentification probabilities, are typically less than 3~\%. 
For this analysis an unfolding procedure
was done by using event weighting with the elements of 
$P^{-1}$ to obtain the hadron multiplicities at the true 
$\pi$ and $K$ fluxes.

In a second step of the analysis 
the particle count rates were corrected for charge
symmetric background processes (e.g. $\gamma \rightarrow e^+ + e^-$).
%
%The rate for this background was estimated by considering lepton
%tracks with a charge opposite to the beam charge that passed the cuts.
%It was assumed that these leptons stemmed from pair-production
%processes and the rate for the charge symmetric background process (where
%the particle is detected with the same charge as the beam but
%originating from pair production) is the same. 
%
The number of events
with an opposite sign lepton is %therefore
an estimate of the number of
charge symmetric events that masquerade as DIS or SIDIS events.
The overall background fraction from this source was 1.4~\%.

The contribution of diffractive processes to the inclusive and
semi-inclusive data were determined with the help of a Monte-Carlo
simulation. This simulation is based on Pythia 6~\cite{pythia}, 
modified by HERMES
to describe the cross section and the angular distribution of
the two meson decay ( e.g. $\rho \rightarrow 2 \pi $ and $\phi
\rightarrow 2 K$) correctly.
It turned out that this contribution is significant for the HERMES
kinematic especially for the charged $\pi$ multiplicities. The
correction is as large as  40~\% for the highest $z$-bins.  

\begin{figure}
\centering
\psfig{figure=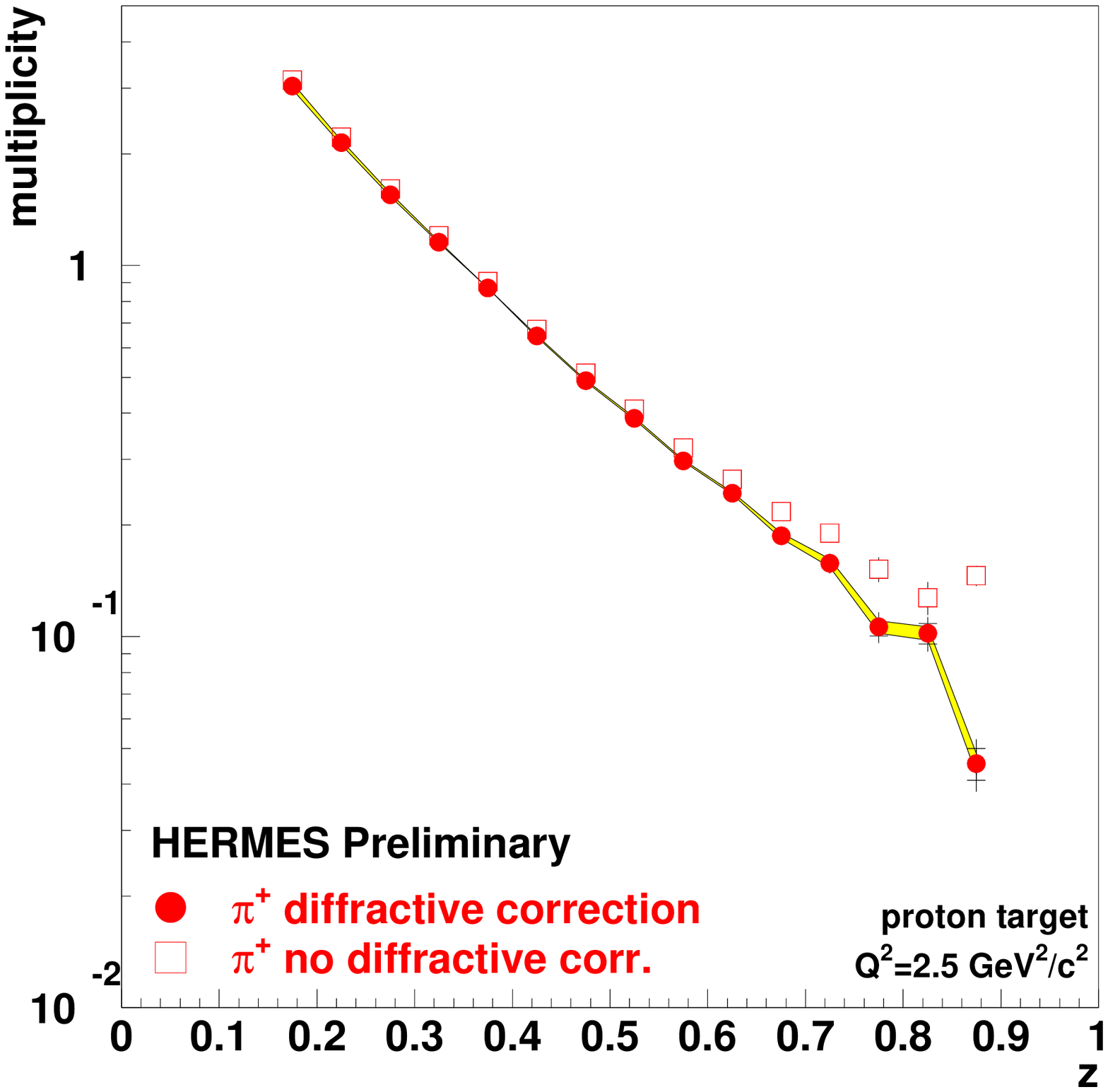,height=7.5cm}
\psfig{figure=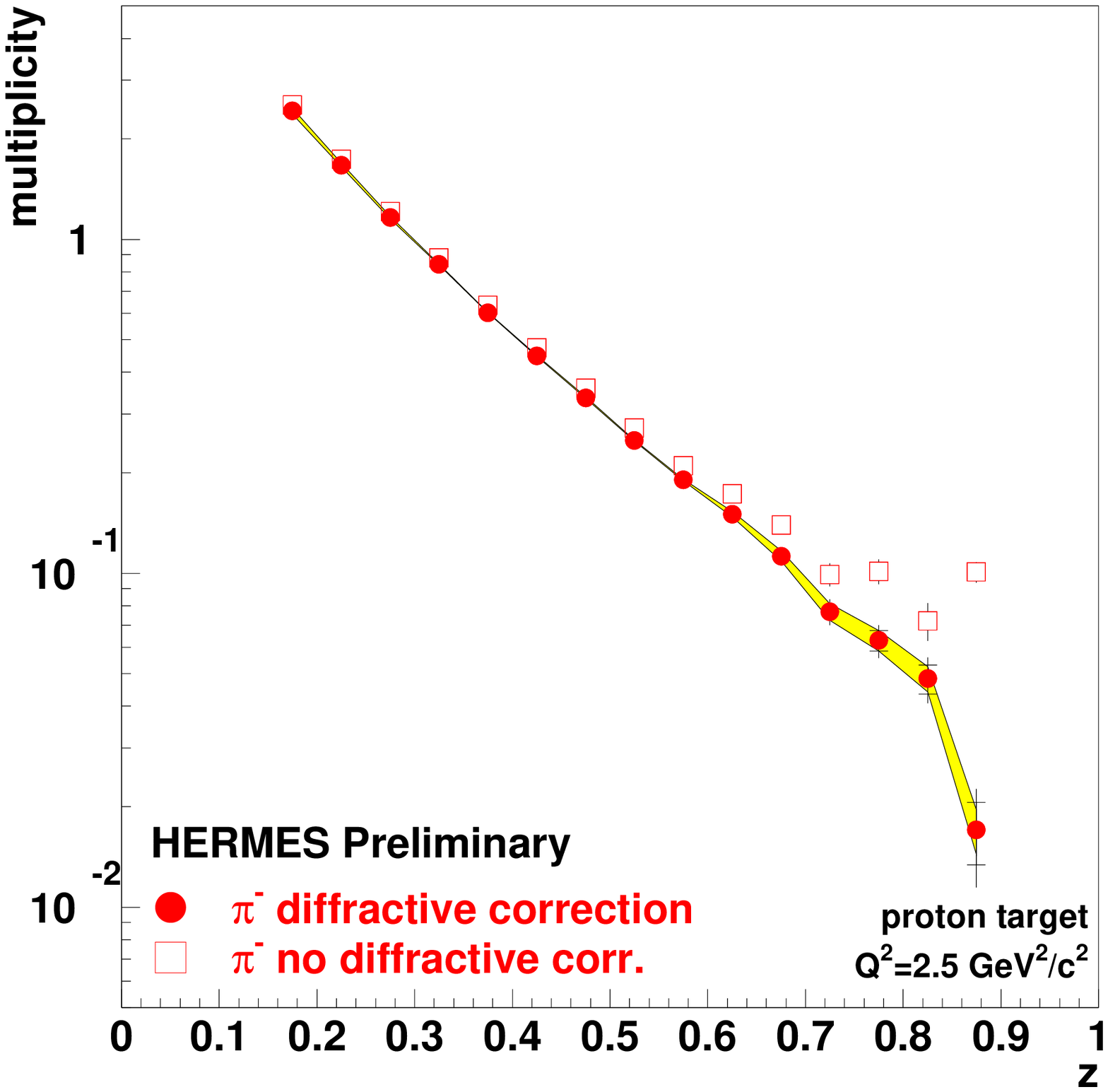,height=7.5cm}
\caption{$\pip$ and $\pim$ multiplicities vs z at
 an average $Q^2$ = 2.5~GeV$^2$. 
\label{fig:pi_mult}}
\centering
\end{figure}

Furthermore the count rates were corrected for detector acceptance and
smearing and QED radiative effects to obtain the Born multiplicities
in 4$\pi$. The corrections were applied using an unfolding algorithm
that accounts for the kinematic migration of the events. A description 
of the unfolding algorithm is found in reference~\cite{delta_q}.

As the HERMES detector has a rather limited acceptance, the average $Q^2$ 
for each $z$-bin changes. The multiplicities were
evolved to a common $Q^2$ = 2.5~GeV$^2$ using a multiplicative 
factor in each bin.
The $Q^2$-evolution was performed using equation (1). 
%\begin{equation}
%  \frac{1}{\sigma_{dis}
%  \frac{d\sigma^{h}(z,Q^{2})}{dz} =
%  \frac{\sum_{f}e_{f}^{2}\int_{0}^{1}dx \;
%          q_{f}(x,Q^{2})D_{f}^{h}(z,Q^{2})}{\int_{0}^{1} \frac{dx}{x}
%          F_2(x,Q^2)}
%\end{equation}
The PKH parameterization~\cite{pkh} for the fragmentation function 
and the CTEQ6 parameterization~\cite{cteq6} for the parton distribution 
functions were used as input for the calculation. The correction factor
is $\approx$ 1~\%. The contribution 
to the systematic error is very small, 2~\% in the lowest
$z$-bin and negligible for the rest of the bins.

\section{Results}

Figure~\ref{fig:pi_mult} shows the extracted positive (left plot) 
and negative (right plot) pion multiplicities as a function of $z$
for a hydrogen target. The data, with very small errors bars, 
show the expected exponential 
decrease. The inner error bars represent the 
statistical error. The error band shows the systematic error
from the RICH misidentification. The error outside the tick marks
is the total systematic error except the one from the RICH.
The differences between the filled symbols and 
the open symbols, which show the multiplicities without diffractive 
correction, demonstrate the impact of this correction especially at
higher $z$-values. This step in the analysis is a significant 
improvement compared to the previous HERMES publication~\cite{pub_mult}. 
The extracted multiplicities evolved to $Q^2$ = 25~GeV$^2$ are in
nice agreement with EMC results~\cite{emc}. 

The charged pion multiplicities for HERMES can also be determined as a
function of $x$ and $Q^2$ for four different $z$-bins. For this
analysis, only data in the range of $0.25 < z < 0.75 $ were
considered. They show only a weak $x$- and $Q^2$-dependence. 

\begin{figure}
\centering
\psfig{figure=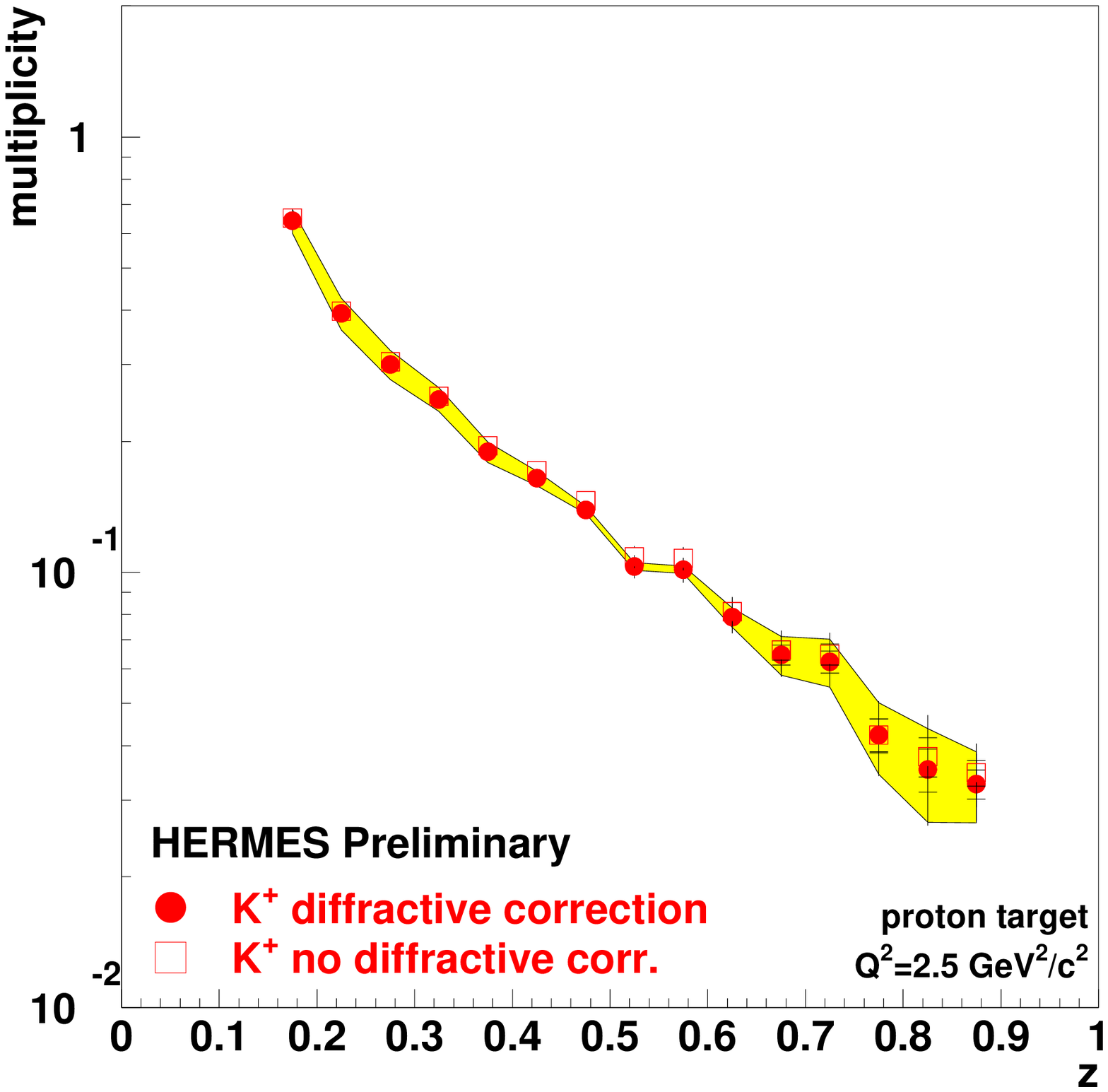,height=7.5cm}
\psfig{figure=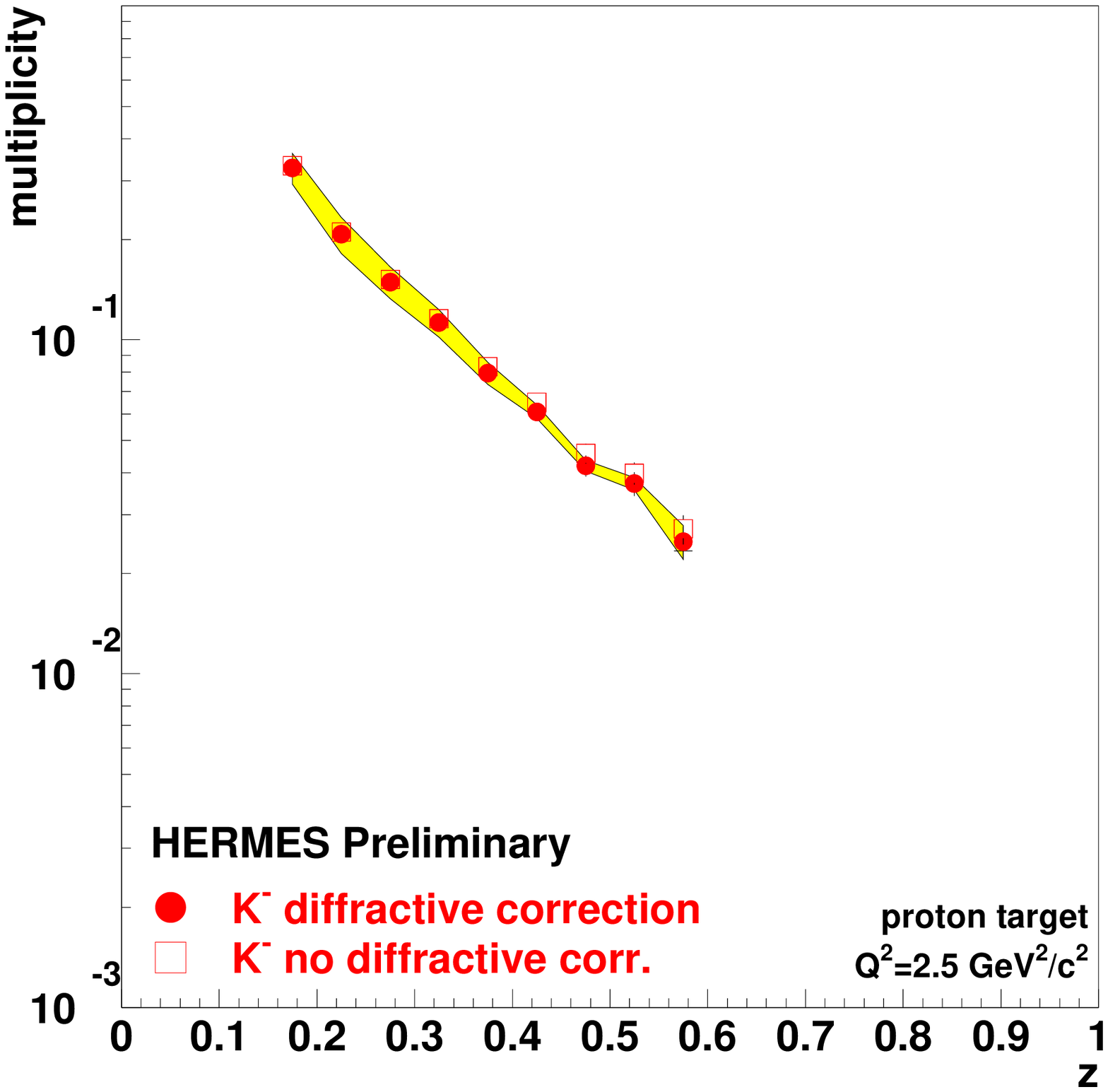,height=7.5cm}
\caption{$\kp$ and $\km$ multiplicities vs z at
 an average $Q^2$ = 2.5~GeV$^2$. 
\label{fig:k_mult}}
\centering
\end{figure}

The charged kaon multiplicities show the same behavior like
for the charged pion multiplicities. They are plotted
in figure~\ref{fig:k_mult} as a function of $z$.  For the $\km$
the range is restricted to $z < 0.6$,  due to the limited
statistics. The errors of the measurements are 
dominated by the systematic error coming from the RICH kaon 
misidentification. 

\section{Outlook}

The charged pion and kaon multiplicities in semi-inclusive scattering
at 27.5~GeV on hydrogen have been measured by the HERMES
collaboration. The measurements provide data in the lower 
$Q^2$-range at an average $Q^2$ = 2.5~GeV$^2$ and in the range
of 0.023 $< x <$ 0.6. 
%The increased 
% taken during the 2000 HERA running period 
%improved the statistical and systematic errors. 
The measured multiplicities will allow the extraction of 
fragmentation functions in the range of 0.15 $< z <$ 0.9. 

HERMES will repeat the extraction of hadron multiplicities 
for the data taken with an unpolarized deuterium target.
A combined analysis of the hydrogen and deuterium multiplicities
will yield favored and unfavored fragmentation functions independent 
of parton distribution functions.

\end{document}